\documentclass[12pt,preprint]{aastex}

\begin{document}


\title{An Unbiassed Census of Active Galactic Nuclei in the Two Micron All Sky
Survey}


\author{Paul J. Francis\altaffilmark{1}}
\affil{Research School of Astronomy and Astrophysics, the Australian
National University, Canberra 0200, Australia}
\email{pfrancis@mso.anu.edu.au}

\altaffiltext{1}{Joint Appointment with the Department of Physics, 
Faculty of Science, the Australian National University}

\and

\author{Brant O. Nelson and Roc M. Cutri}
\affil{Infrared Processing and Analysis Center, California Institute of
Technology, Mail Code 100-22, 770 South Wilson Avenue, Pasadena, CA 91125}
\email{nelson, roc@ipac.caltech.edu}

\begin{abstract}

We present an unbiassed near-IR selected AGN sample, covering 12.56
square degrees down to $K_s \sim 15.5$, selected from
the Two Micron All Sky Survey (2MASS). Our only selection effect is a
moderate color cut ($J-K_s>1.2$) designed to reduce contamination from galactic
stars. We observed both point-like and extended sources.
Using the brute-force capabilities of the 2dF multi-fiber spectrograph on the
Anglo-Australian Telescope, we obtained spectra of 65\% of the target list:
an unbiassed sub-sample of 1526 sources. 

80\% of the 2MASS sources in our fields are galaxies, with a median redshift
of 0.15. The remainder are K- and M-dwarf stars. 

We find tentative evidence that Seyfert-2 nuclei are more common in our
IR-selected survey than in blue-selected galaxy surveys. We estimate
that $5.1 \pm 0.7$\% of the galaxies have Seyfert-2 nuclei with H$\alpha$
equivalent widths $> 0.4$nm, measured over a spectroscopic aperture of radius 
$\sim 2.5$kpc. Blue selected galaxy samples only find Seyfert-2 nuclei
meeting these criteria in $\sim 1.5$\% of galaxies.

$1.2 \pm 0.3$\% of our sources are broad-line (Type-1) AGNs, giving a 
surface density of $1.0 \pm 0.3$ per square degree, down to $K_s <15.0$. 
This is the same surface density of Type-1 AGNs as optical samples down 
to $B<18.5$. Our Type-1 AGNs, however, mostly lie at low redshifts, and host
galaxy light contamination would make $\sim 50$\% of them hard to find in 
optical QSO samples.

We conclude that the Type-1 AGN population found in the near-IR is not 
dramatically different from that found in optical samples. There is no evidence for
a large population of AGNs that could not be found at optical wavelengths,
though we can only place very weak constraints on any population of dusty
high-redshift QSOs. In contrast, the incidence of Type-2 (narrow-line)
AGNs in a near-IR selected
galaxy sample seems to be higher than in a blue selected galaxy sample.

\end{abstract}

\keywords{galaxies: active --- Surveys --- 
quasars: general}

\section{Introduction}

To date, nearly all complete Active Galactic Nuclei (AGN) samples are
flux limited at blue optical wavelengths. Such surveys are highly efficient, 
and can be very complete \citep[eg.][]{mey01}, picking up all AGNs {\em 
down to their blue flux limit}. Unfortunately, any survey with a blue flux 
limit will be relatively insensitive to objects whose emission peaks at any 
other wavelength. 

How seriously does this blue flux limit bias AGN samples? The situation is
somewhat different for QSO searches (searches for AGNs which are considerably
brighter than their host galaxy) and Seyfert galaxy searches (searches for less
luminous AGNs).

\subsection{Luminous QSOs}

\begin{figure}
\plotone{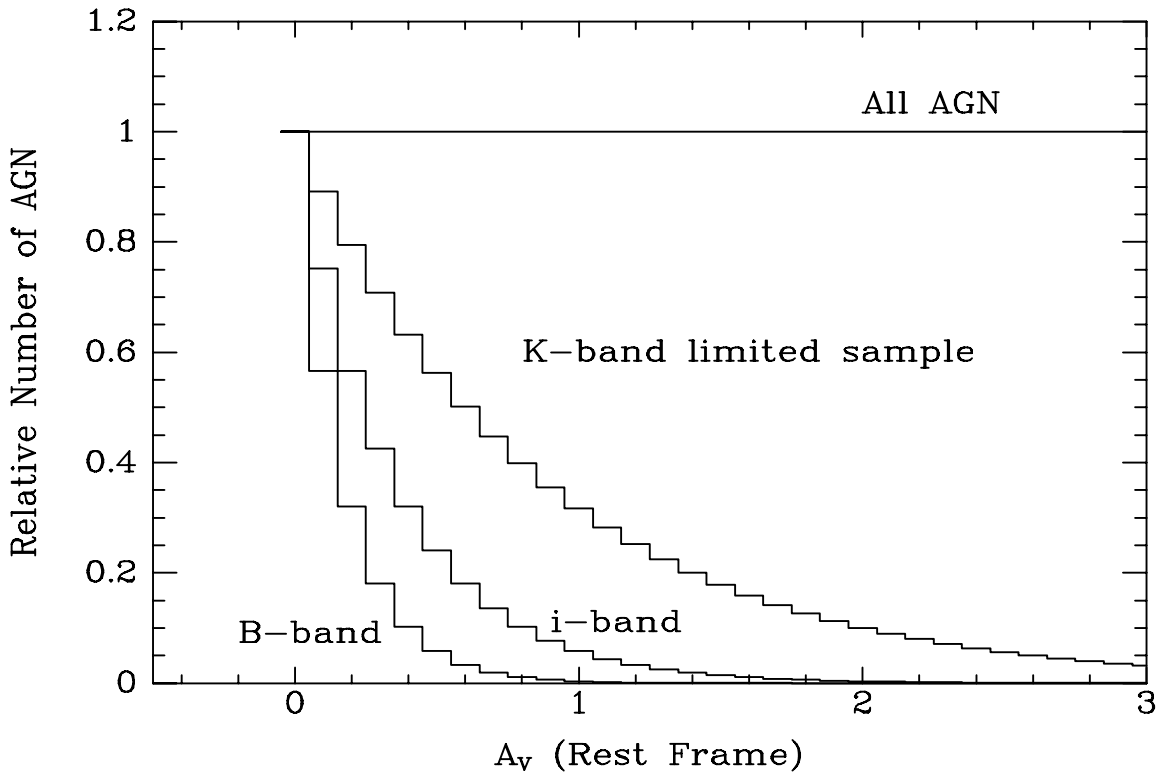}
\caption{
Predicted numbers of QSOs found as a function of dust extinction. The model
assumes that the real population of QSOs is uniformly distributed per unit
dust extinction $A_V$, where $A_V$ is the absorption in the rest-frame 
$V$-band, in magnitudes (the QSOs are assumed to lie at redshift 
one). The labeled curves show the fraction of these QSOs that would
be found in complete surveys, magnitude limited in the $B$, $i$ and
$K_s$ bands. A luminosity function appropriate for bright QSO samples
has been assumed. Dust is assumed to have an optical depth inversely
proportional to wavelength, and to lie at the QSO redshift.
\label{dust}}
\end{figure}

There has long been speculation that there might exist a substantial
population of luminous QSOs with red colors in the optical/near-IR.
These red colors could be caused by small quantities of dust, or 
the QSOs could be intrinsically red. Given
the steepness of the luminosity function for luminous QSOs, most 
will lie close to the magnitude limit of a survey, so even small amounts of
extinction will eliminate them from a blue-selected sample (Fig~\ref{dust}). 

QSOs live in the nuclei of galaxies, which are dusty places. It should 
therefore be no surprise that our sight-line to the centers of many QSOs is 
obscured by dust. What is surprising is that the dust seems to either 
completely obscure our view of the central engine of the QSO (Type-2 AGN), 
or not to obscure it at all (Type-1 AGN). There seem to be very few QSOs that 
are partially obscured by dust, so that we still see a nuclear QSO spectrum, 
albeit a reddened one. Our sight-line seems either to intersects a giant 
molecular cloud or
no dust at all. This contrasts with sight-lines from the Earth out of
our galaxy, most of which intersect small quantities of optically thin dust
\citep{sch98}. Is this a selection effect, or does AGN activity expel or 
destroy optically thin dust, as suggested by \citet{dop98}? 

A few red QSOs have now
been found. Many radio-selected quasars are quite red, though this redness
may be caused by synchrotron emission or weak blue bump
emission, rather than dust \citep{web95,bak95,whi01,fra00,fra01}. At least
a few radio-selected quasars, however, show unmistakable evidence of severe
dust reddening \citep{mal97,cou98,gre02}. A handful of red AGNs
have also been identified in other surveys \citep[eg.][]{mcd89,bro98}.

To accurately determine the population of red QSOs, and to better characterize
their apparently diverse nature, a QSO sample with a magnitude limit at some
wavelength unaffected by dust would be ideal. Radio surveys only pick up
the small fraction of QSOs that are radio-loud, which are probably not
representative. Hard X-ray surveys \citep[eg.][]{mus00,ale01} are unaffected
by dust, but many hard X-ray sources are so faint at optical wavelength that
follow-up spectroscopy is very difficult, even with large telescopes. It
does, however, seem clear that dusty AGNs are a major contributer to the
X-ray background. Far-IR selection \citep[eg.][]{low88,mat02} is biased
{\em towards} dusty sources, but discriminating between QSOs and starburst
galaxies has proven extremely hard.

Complete near-IR selected surveys are still somewhat biased against dusty QSOs
(Fig~\ref{dust}). They have the major advantage that most QSOs found in a
near-IR limited survey will be bright enough for relatively easy follow-up
spectroscopy. Surveys with $i$-band magnitude limits, such as the
Sloan Digital Sky Survey (SDSS) QSO survey \citep{ric02}, are an
improvement on $B$-band limited surveys, but Fig~\ref{dust} makes it
clear that going still further to the red should yield big gains.

Can we construct a complete $K$-band limited QSO sample?
\citet{war00} showed that by combining optical and near-IR photometry, it
should be possible to construct such a sample. Unfortunately, suitable
photometry does not yet exist over larger areas of the sky, though the
technique has been successfully applied in one small region \citep{cro01}.

\subsection{Seyfert Nuclei}

The situation is somewhat different for less luminous AGNs. These cannot
be found by color selection, as the host galaxy light dominates their
broad-band colors. They are normally found by getting spectra of the nuclear
regions of large samples of galaxies \citep[eg.][]{huc92,ho97}. To date,
these galaxy samples have been magnitude limited in the blue. This may well
introduce a bias: the blue light from galaxies is dominated by young stars,
and is hence an indication of recent star formation.

The near-IR light from galaxies is coming from an older stellar population,
and hence correlates with the total stellar mass rather than the recent 
star formation rate. Near-IR selected galaxy samples are dominated by
elliptical galaxies, unlike blue selected samples which are
dominated by spirals.

We might thus expect the population of AGNs in an IR-selected galaxy
sample to differ from that in a blue-selected sample for many reasons.
The black hole masses, which are known to correlate with the bulge stellar 
mass, should be larger. If accretion onto the black hole correlates
with star formation, we might be looking at lower accretion rates. Dust
properties may be quite different, altering the ratios of obscured (Type-2) 
and unobscured (Type-1) AGNs.

\subsection{Searching for AGNs in 2MASS\label{search}}

By far the largest near-IR survey to date is the Two Micron 
All Sky Survey \citep[2MASS, ][]{skr97}. There
have already been several studies of the different AGN populations within 
2MASS. \citet{cut02} have shown that 2MASS sources with extremely red
near-IR colors ($J-K_s>2$) are mostly an unusual type of Type-1 AGN
\citep{smi02,wil02}. $K_s$ is a filter similar to $K$
but cutting off at a shorter red wavelength to minimize thermal emission
\citep{skr97}. \citet{bar01} have studied the 2MASS colors of QSOs identified
at other wavelengths, and \citet{gre02} identified some very unusual and
dusty QSOs by cross-correlating the 2MASS database with a radio sample.
While these various papers clearly show that 2MASS imaged large numbers
of AGNs, none of them made any pretense at giving an unbiassed picture
of the AGN population within 2MASS.

In this paper, we assemble a relatively unbiassed sample of 2MASS
AGNs. We use brute force: we apply only a very weak color selection,
and then use multi-object spectroscopy to pick out the few AGNs from the
large contamination of other objects.

\begin{figure}
\plotone{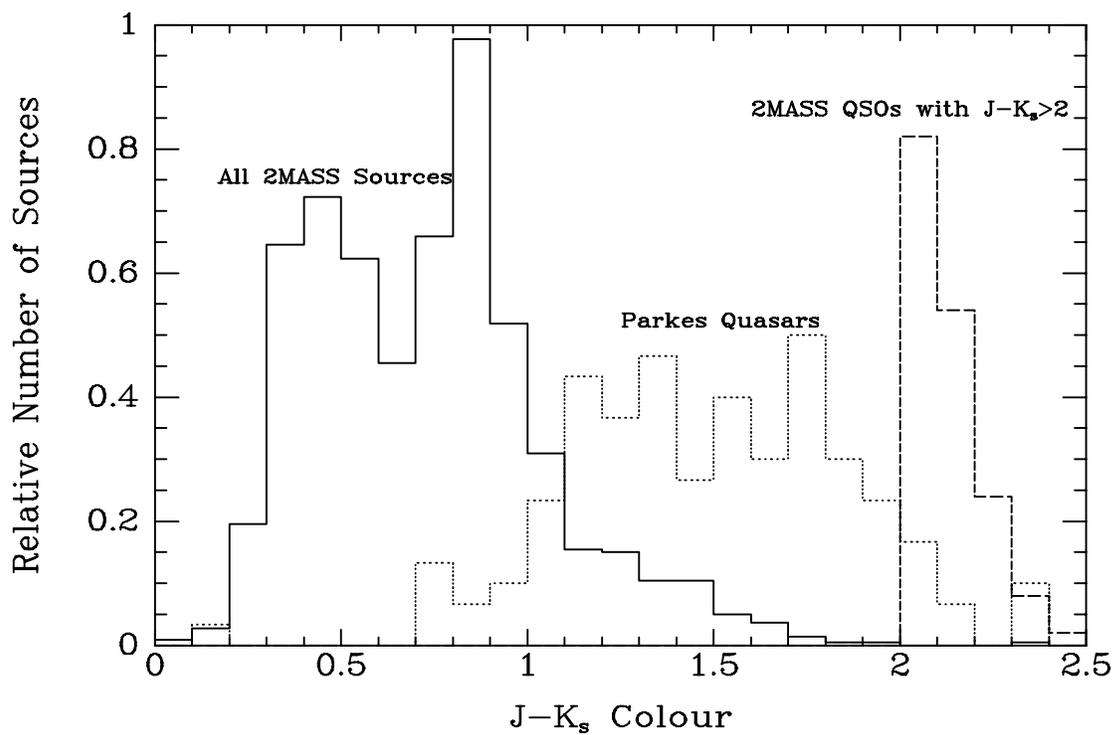}
\caption{The distribution of $J-K_s$ colors for high galactic latitude 
2MASS sources (solid
line), Type-1 AGN selected as having $J-K_s>2$ (dashed line. Cutri et al.), 
and radio selected quasars from the Parkes Half Jansky Flat Spectrum sample 
\citep[dotted line,][]{fra00}
\label{jkhist}}
\end{figure}

Our only selection criteria were that an object had to be detected in all
three 2MASS bands, and that it had $J-K_s>1.2$. This color cut was
designed to eliminate halo giant and disk dwarf stars (the peaks at 
$J-K_s \sim 0.45$, $0.75$ in Fig~\ref{jkhist}), but still be sensitive to
most galaxies and QSOs. Nearly all QSOs with redshifts below $\sim 0.5$,
selected at other wavelengths, have $J-K_s>1.2$ \citep{fra00,bar01,cut02}.
At higher redshifts, the near-IR flux excess \citep{san89} is shifted out 
of the $K_s$-band, causing the average $J-K_s$ color of known AGN to become
bluer, but even at these higher redshifts, at least 10\% of AGN have
$J-K_s>1.2$.

The color cut eliminates some galaxies from our sample.  The median
$J-K_s$ color of galaxies in the 2MASS Extended Source Catalog (XSC) is 
$\sim 1.1$.
The mean color of galaxies shifts rapidly to the red with increasing
redshift due to k-corrections, though.  At the magnitude limit of
the 2MASS Point Source Catalogue (our input catalog), most galaxies will 
be unresolved and will lie
at redshifts well above 0.1 where the median galaxy color is redder
than $J-K_s = 1.2$.  We estimate our incompleteness to galaxies by
counting the number of 2MASS XSC sources with $J-K_s < 1.2$ in
our survey areas, and by cross-correlating the Sloan Digital Sky Survey
Early Release galaxy catalog with the 2MASS PSC:  fewer than 23\%
of galaxies would fail to meet our color cut.  The missing
galaxies will be predominantly at redshifts less than 0.1
and resolved by 2MASS.

The bias of our survey towards low redshift AGN does limit our ability to
find dusty QSOs. The luminosity of most QSOs found in the local universe is
only a little greater than that of their host galaxies. Even small amounts of
dust extinction will thus reduce the AGN light below the host galaxy light,
causing the source to be classified as a Type-2 AGN rather than a dusty
Type-I AGN. Spectacularly reddened Type-1 AGN should thus only be found in
high redshift, high luminosity samples, such as that of \citet{gre02}.

Our target selection, observations and data reduction are described in
\S~\ref{obsred}, and the spectral classification of our
sources in \S~\ref{class}. Our results are presented in
\S~\ref{results} and discussed in \S~\ref{discuss}. Finally,
conclusions are drawn in \S~\ref{conclude}

\section{Observations and Reduction\label{obsred}}

\subsection{Target Selection}

Targets were selected from the 2MASS Point Source Catalog.
Note that this catalog includes extended sources.
All cataloged sources with $J-K_s>1.2$ and detections in all three bands
were potential targets, regardless of
optical magnitude or morphology. No attempt was made to exclude previously 
observed sources. 

We observed spectra of sources in four fields. Each field was circular, and 
one degree in radius. The fields were centered at 
09:44$+$00:00, 12:44$+$00:00, 13:00$-$25:00 and 14:15$-$26:00 (J2000).
The first two fields were chosen to overlap with the imaging data from the
early data release of the Sloan Digital Sky Survey \citep[SDSS,][]{sto02}.
All fields lie at galactic latitudes greater than 30 degrees.

\begin{figure}
\plotone{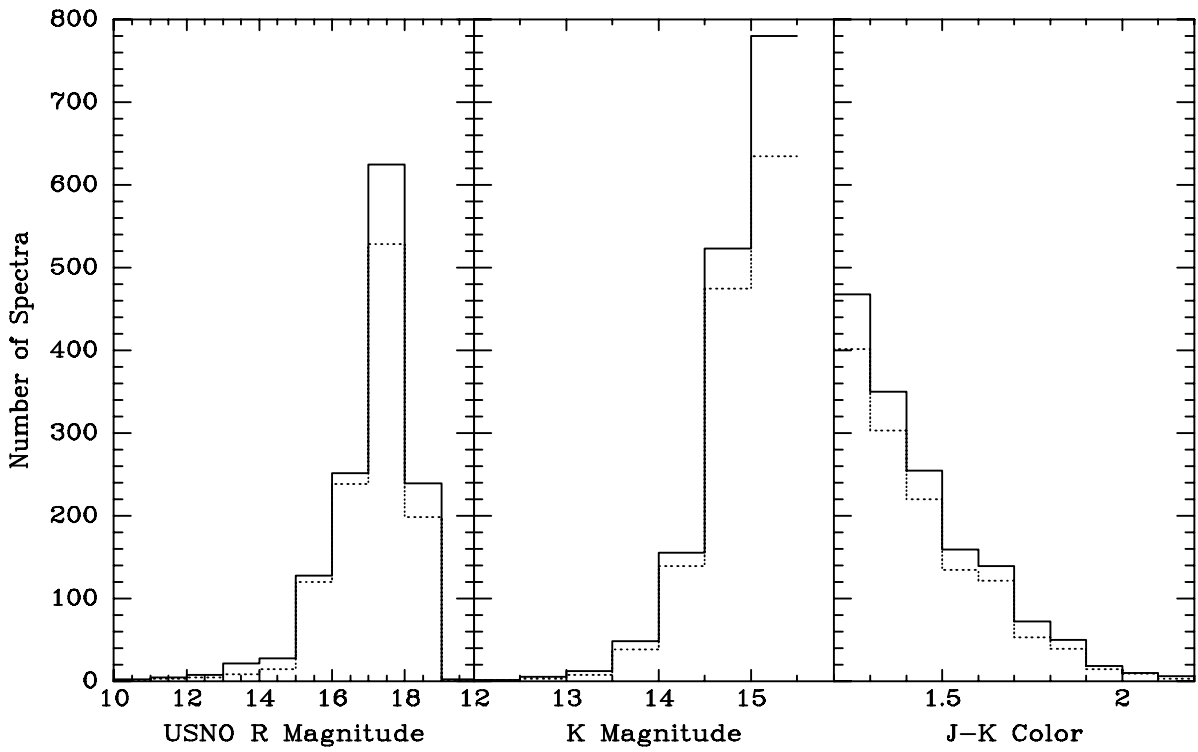}
\caption{
The number of spectra observed (solid line) and the number for which
we secured a reliable spectral identification, either manually or automatically
(dotted line), as a function
of R-magnitude (taken from the USNO-A catalog), $K_s$-magnitude and $J-K_s$ color.
\label{complete}}
\end{figure}

Observations were carried out with the Two Degree Field (2dF) spectrograph on
the Anglo-Australian Telescope \citep[AAT,][]{lew02}. This spectrograph
has 400 fibers, spread over a circular field of radius one degree, located at
the prime focus of the AAT. Each fiber has a projected diameter of 2\arcsec\ 
on the sky. A small number of fibers were set aside to measure
the sky spectrum. The remaining
fibers were allocated to targets using the {\it configure} program
\citep[][]{lew02}. The program was set to allocate fibers to the brightest
$K_s$-band sources first, and then progressively to the fainter ones.
We were able to allocate fibers to all 69 sources
with $K_s <14.0$, 677 of the 873 sources with $14.0 < K_s < 15.0$, but only
780 of the 1407 sources with $15.0 < K_s < 15.5$. The incompleteness in the
$14.0 < K_s < 15.0$ range is mostly due to fiber positioning constraints, while
the incompleteness at fainter magnitudes is due to the limited number of fibers.
The incompleteness that
this introduces should be random in every parameter except K-band magnitude.
The magnitude and color distribution of the sources for which we obtained
spectra are shown in Fig~\ref{complete}.

\subsection{Observations and Reduction}

Spectra were taken of sources in our four fields on the nights of 2002 
March 5 -- 7. Conditions were partially
cloudy at times, and the seeing was typically around 1.8\arcsec . Each field
was observed with two different fiber configurations: one for the brightest
$\sim 30$ sources and the other for the remaining $\sim 350$. This technique
was chosen to minimize scattered light problems. Exposure times were
600-900 sec for the bright object configurations, and 10,036 -- 10,800 sec 
for the
faint object configurations. Bright sources in the 1415$-$2600 field were
not observed, due to cloud. The 300R and 316R gratings were used in the two
spectrographs, giving a spectral resolution of 10\AA\ and a wavelength coverage
of 4500 \-- \ 8500\AA . 

The data were reduced using the {\it 2dfdr} software \citep[][]{lew02}, using
standard settings.
All galaxy spectra were averaged (in the observed frame) to
provide a template atmospheric absorption spectrum. The individual 
spectra were divided
through by this template, which did a reasonably good job of correcting for
these absorption bands. The spectra are not of spectrophotometric quality.

Spectra were obtained for a total of 1526 sources. 59 of these spectra were of
such poor quality that regardless of the nature of the source,
no spectral classification was possible. Fig~\ref{complete} shows that the
unusable quality spectra are predominantly those of the brighter sources.
This was mainly caused by poor weather: the bright objects in one of the four
fields were never observed while in two other fields they were observed 
through significant cloud cover. Observations of the fainter sources were
not as badly affected.

\section{Classification\label{class}}

An initial classification was attempted for all spectra using the 
software developed for the 2dF Galaxy Redshift Survey \citep[][]{col01}.
This software is optimized for measuring galaxy redshifts from 2dF spectra
of comparable quality to our own, and uses template fitting, line-fitting and
cross-correlation techniques to classify spectra and to measure redshifts. All
classifications were checked manually and assigned a quality flag. The program
produced high quality classifications and (where relevant) redshifts for 
around 80\% of our spectra. It showed excellent performance for galaxies in
our sample, but was less reliable for broad-line AGN and stars.
The remaining spectra were checked by eye, and in many cases secure
identifications could be made interactively.

\subsection{Emission line diagnostics\label{eline}}

\begin{deluxetable}{lcc}
\tablecaption{Wavelength Regions Used in Emission-Line Measurements
\label{linewave}}
\tablecolumns{3}
\tablehead{
\colhead{Line} &
\colhead{Continuum Integration Limits (nm)} &
\colhead{Line Integration Limits (nm)}
}
\startdata
H$\beta$     & 474.0 -- 484.0, 488.0 -- 494.0 & 484.5 -- 488.0 \\
$[$\ion{O}{3}$]$   & 488.0 -- 494.0, 503.0 -- 510.0 & 499.0 -- 502.5 \\
H$\alpha$    & 640.0 -- 652.0, 663.0 -- 670.0 & 655.0 -- 657.2 \\
\ion{N}{2}   & 640.0 -- 652.0, 663.0 -- 670.0 & 657.2 -- 659.4 \\
$[$\ion{S}{2}$]$   & 663.0 -- 670.0, 674.5 -- 684.5 & 670.0 -- 674.5 \\
\enddata
\end{deluxetable}

Galaxies showing H$\alpha$ and/or H$\beta$ emission lines with velocity widths
(full width at half maximum height: FWHM) greater than 1000${\rm km\ s}^{-1}$
were classified as Type-1 AGNs. Emission-line ratios were measured 
automatically for the remaining galaxies, by
interpolating a straight-line continuum under them and summing the flux above
this continuum. Wavelength regions used to define the continuum and over which
line fluxes were summed are shown in Table~\ref{linewave}. The effect of the
underlying stellar absorption lines was corrected for by measuring the
mean absorption-line equivalent width of the line-less galaxies in the
sample, and adding this to the measured emission-line equivalent widths.
This assumes that the underlying stellar continua of emission-line and
non-emission-line galaxies are the same, which will only be true to first
approximation.
These corrections are small: 0.05 nm for H$\alpha$, 0.18 nm for H$\beta$,
0.15nm for [\ion{O}{3}, -0.07nm for \ion{N}{2} and 0.005 nm for \ion{S}{2}.
To check that these corrections did not significantly affect our results,
we repeated our classification without making them. This did not alter the
classification of any of our galaxies, principally because the most affected
sources were Seyfert-2 galaxies with very weak H$\beta$ emission, and these
lie a long way from the selection boundary. All line
measurements were checked by eye and awarded a quality flag: 6\% of spectra
were too poor at the relevant wavelength to obtain a good measurement of
H$\alpha$. This was usually caused by H$\alpha$ falling on a strong sky line
or atmospheric absorption band.

\begin{figure}
\plotone{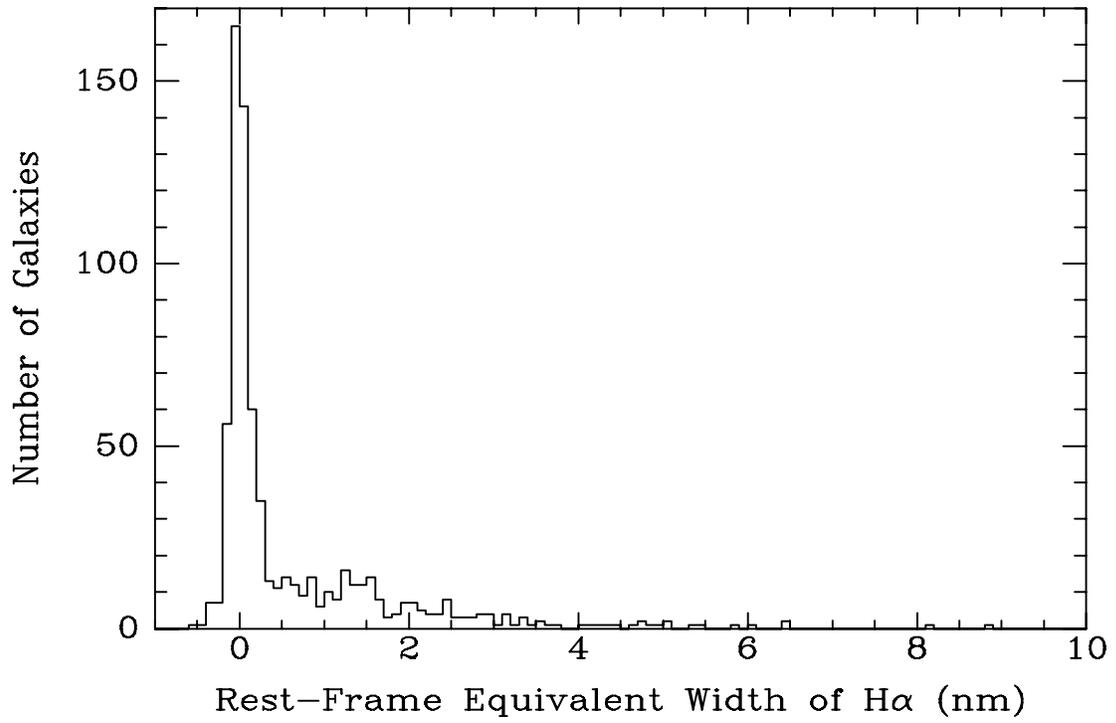}
\caption{Histogram of measured rest-frame H$\alpha$ emission-line
equivalent widths for all galaxies with acceptable quality spectra at the
relevent wavelengths.
\label{hahist}}
\end{figure}

We estimate our equivalent width limit by looking at the dispersion in
H$\alpha$ equivalent width measurements in galaxies with no detectable
line emission (Fig~\ref{hahist}). We estimate that we are sensitive to
all galaxies with rest-frame H$\alpha$ equivalent widths of $>0.4$nm.
This excludes the 6\% of galaxy spectra which were too poor at the relevant
wavelength.

\begin{figure}
\plotone{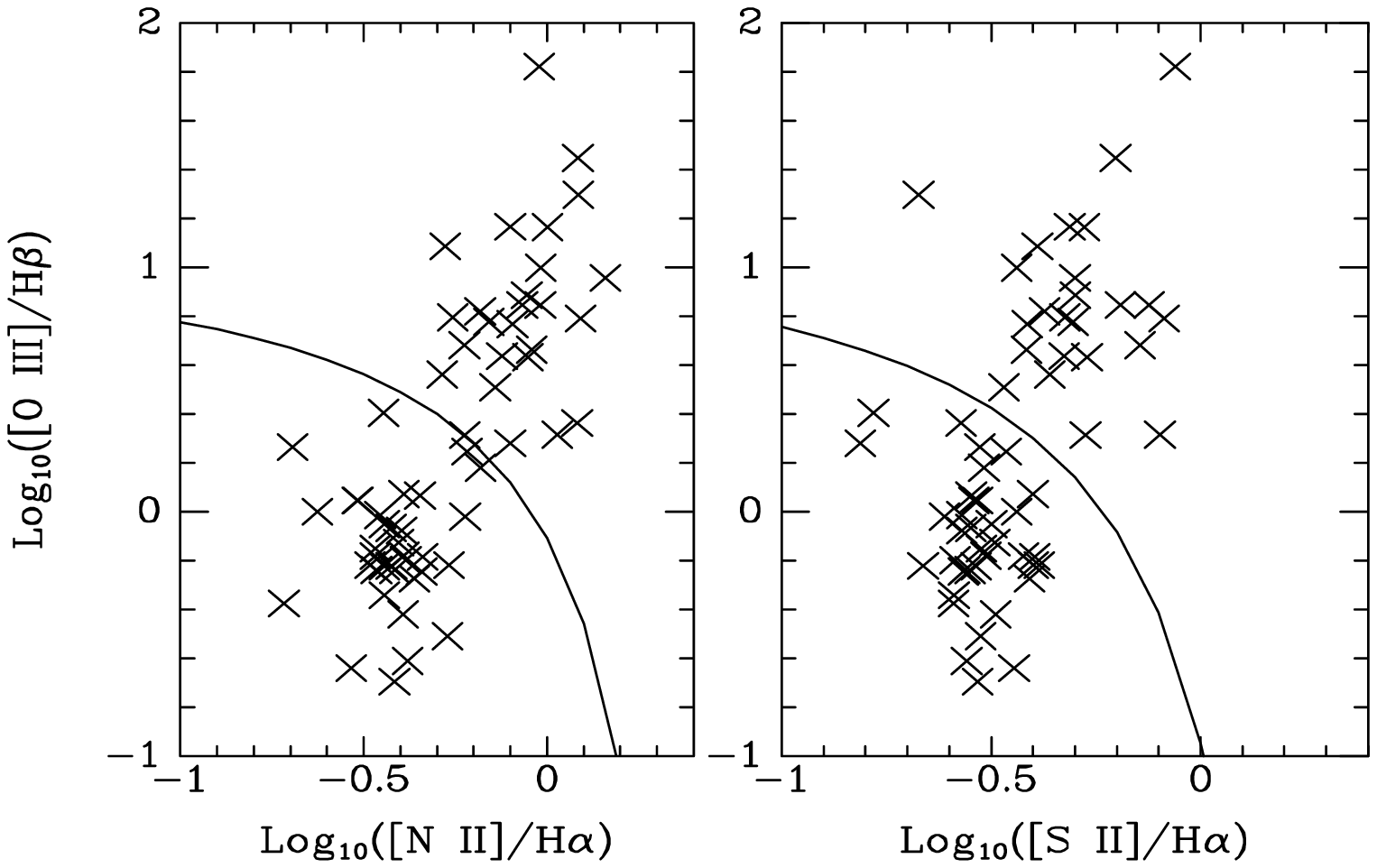}
\caption{
Line ratios of all galaxies with adequate quality data. The solid line
is the theoretical classification boundary from \citet{kew01}. Sources
lying above the boundary have emission-lines excited by an AGN, while
those below the boundary have lines excited by massive stars. One sigma
error bars are typically $\sim 0.1$ in the log. 
\label{classplot}}
\end{figure}

All galaxies with adequate quality data in all emission lines were
classified using the diagnostic diagrams of \citet{kew01}. The results are
shown in Fig~\ref{classplot}. The galaxies split cleanly between AGN and
starbursts. The AGN have the line ratios of Seyfert-II galaxies and not
of LINERS. The 23 sources lying above both classification lines were
classified as Type II AGN, and the 34 lying below as Starburst galaxies.
One source lay above one line and below the other: we classified it as
an unknown emission-line galaxy.

Unfortunately, while many galaxies had good quality data for the lines
near H$\alpha$ (\ion{N}{2} and $[$\ion{S}{2}$]$), the shorter wavelength 
lines (H$\beta$ and $[$\ion{O}{3}$]$) were often too weak for us to
be able to calculate their position along the y-axis of Fig~\ref{classplot}.
We note, however, a reasonably strong correlation between the x- and
y-axes in the classification plots. If this correlation holds for the galaxies
with weaker short wavelength lines, we can use it to tentatively classify
at least some of these sources. All otherwise unclassified emission-line
galaxies with $\log_{10}($\ion{N}{2}$ / H \alpha ) > -0.2$ and
$\log_{10}([$\ion{S}{2}$]/ H \alpha ) > -0.35$ were classified as
probable AGN, while sources with  $\log_{10}($\ion{N}{2}$ / H \alpha ) < -0.3$ and
$\log_{10}([$\ion{S}{2}$] / H \alpha ) < -0.4$ were classified as probable
starbursts. This yielded another 12 probable Seyfert II galaxies, and
65 probable starburst galaxies. All other galaxies with H$\alpha$ equivalent
widths greater than 0.4nm were classified as unknown emission-line galaxies.

\subsection{Sloan Digital Sky Survey Data\label{sdss}}

We extracted data from the SDSS early data release for the 739 of our
targets lying within its region of coverage. For all but two of these 
sources, a SDSS cataloged source was found within 1.6\arcsec of the
2MASS position. The median positional offset between the 2MASS and SDSS
coordinates was 0.269\arcsec. 

Two 2MASS sources did not have SDSS cataloged sources within 5\arcsec\
of the 2MASS position: 2MASS 0943007$-$000955 (an M-dwarf star) and
2MASS 0946501$+$002050 (a galaxy at redshift 0.1414). As we obtained
good spectra of both sources, using the 2.0\arcsec\ diameter 2dF fibers
centered at the 2MASS coordinates, the error must not lie in the 2MASS 
catalog.

All the 477 sources we classified as galaxies (on the basis of their 
spectra) were classified as extended sources by SDSS, while 147/156
stars were classified as point sources. Of the 68 sources with SDSS
data that we were unable to classify based on their spectra, 54 (80\%) were
classified by SDSS as extended sources. A visual inspection of their
spectra confirms that most are probably galaxies without strong emission or
absorption lines at wavelengths with good data.

\section{Results\label{results}}

We obtained 1526 spectra, or which 1467 were of usable quality. We were
able to obtain secure classifications for 1298 of these spectra (88\% 
completeness). As noted in Section~\ref{sdss}, the remaining unclassified
sources are predominantly galaxies without strong absorption or emission
lines at wavelengths for which we have good data. If any of these
unclassified sources with usable quality spectra were AGNs with H$\alpha$
equivalent widths of $> 0.4$nm, we would 
have detected their emission lines. The unclassified sources are concentrated
in the fields observed through cloud: our successful classification rate is
much higher in the fields observed in clear weather.

\begin{deluxetable}{cccccc}
\tabletypesize{\scriptsize}
\tablecaption{Source classifications as a function of J-K color \label{tclass}}
\tablehead{
\colhead{$J-K_s$} & 
\colhead{Number of Classified Sources} & 
\colhead{Stars} & 
\colhead{Galaxies} & 
\colhead{Type 1 AGNs} & 
\colhead{Type 2 AGNs}
}
\startdata
\sidehead{Data in this paper}
1.2 -- 1.4 & 707 & 33\% & 67\% & 0.3\% & 1.7\% \\
1.4 -- 1.6 & 357 & 19\% & 81\% & 1.4\% & 2.5\% \\
1.6 -- 1.8 & 174 & 15\% & 85\% & 1.7\% & 1.7\% \\
$>$ 1.8    &  66 &  9\% & 91\% & 6.0\% & 0.0\% \\
\sidehead{Data from Cutri et al. (2003)}
$>$ 2.0    & 664 & 1\% & 99\% & 58\% & 15\% \\
\enddata
\end{deluxetable}

330 (25\%) of the classified objects are stars. Around 20\% are late 
K-dwarfs, and the remainder are M-dwarfs. The stars with SDSS data have a
median $R=19.22$. The fractions of objects with various classifications 
as a function of $J-K_s$ color are shown in Table~\ref{tclass}.

\begin{deluxetable}{lcccccl}
\tabletypesize{\scriptsize}
\tablecaption{Type-1 AGNs \label{type1}}
\tablehead{
\colhead{Name} & 
\colhead{Position (J2000)} & 
\colhead{R} & 
\colhead{$K_s$} & 
\colhead{$J-K_s$} & 
\colhead{Redshift} & 
\colhead{Previous Name}
}
\startdata
2MASS 09403186-0028433 & 09:40:31.86 -00:28:43.3 & 18.2 & 15.38 & 1.39 &
0.153 & \nodata \\
2MASS 09441580+0011015 & 09:44:15.80 +00:11:01.5 & 17.2 & 14.61 & 1.51 & 
0.128 & sdss J094415.78+001101.2 \\
2MASS 09452492+0041448 & 09:45:24.92 +00:41:44.8 & 17.8 & 15.04 & 1.65 & 
0.200 & \nodata \\
2MASS 09460212+0035186 & 09:46:02.12 +00:35:18.6 & 18.1 & 14.85 & 1.73 & 
0.649 & sdss J094602.11+003518.7 \\
2MASS 12420264+0012191 & 12:42:02.64 +00:12:29.1 & 16.9 & 14.63 & 1.46 & 
1.216 & LBQS 1239+0028 \\
2MASS 12442311+0027160 & 12:44:23.11 +00:27:16.0 & 17.8 & 15.02 & 1.49 & 
0.165 & sdss J124423.07+002715.9 \\
2MASS 12452459-0009379 & 12:45:24.59 -00:09:37.9 & 17.6 & 15.42 & 1.22 & 
2.077 & LBQS 1242+0006 \\
2MASS 12461313-0042330 & 12:46:13.13 -00:42:33.0 & 16.7 & 14.46 & 1.49 & 
0.649 & LBQS 1243-0026 \\
2MASS 13025113-2428552 & 13:02:51.13 -24:28:55.2 & 17.5 & 14.50 & 2.17 & 
0.246 & \nodata \\
2MASS 13031854-2435071 & 13:03:18.54 -24:35:01.7 & 17.4 & 14.69 & 1.84 & 
2.255 & HB1300-243 \\
2MASS 14161423-2607468 & 14:16:14.23 -26:07:46.8 & 17.5 & 14.25 & 1.70 & 
0.220 & \nodata \\
2MASS 14165581-2524134 & 14:16:55.81 -25:24:13.4 & 15.3 & 12.73 & 1.82 & 
0.236 & CTS0025 \\
2MASS 14180763-2548430 & 14:18:07.63 -25:48:43.0 & 16.8 & 14.73 & 2.02 & 
0.494 & \nodata \\
2MASS 14183782-2540138 & 14:18:37.82 -25:40:13.8 & 16.8 & 14.55 & 1.48 & 
0.155 & \nodata \\
\enddata
\end{deluxetable}

The remaining 968 sources (75\%) are galaxies of various types. 14 have 
broad emission lines and are hence Type-1 AGN (Table~\ref{type1}). 23
are definite Type-2 AGNs (Seyfert 2 galaxies) while another 12 are
probable Type-2 AGNs (as described in Section~\ref{eline}).
The Type-2 AGNs are listed in Table~\ref{type2}. There are 106 galaxies
whose line ratios make them definite or probable starburst galaxies, and
a further 71 galaxies with H$\alpha$ rest-frame equivalent widths greater 
than 0.4nm , but which we couldn't classify. The galaxies with SDSS data have a
median $R=17.65$. Redshifts and spectral classifications for all galaxies with
adequate data are shown in Table~\ref{allgal}.

\begin{deluxetable}{lcccccl}
\tabletypesize{\scriptsize}
\tablecaption{Type-2 AGNs \label{type2}}
\tablehead{
\colhead{Name} & 
\colhead{Position (J2000)} & 
\colhead{R} & 
\colhead{$K_s$} & 
\colhead{$J-K_s$} & 
\colhead{Redshift} & 
\colhead{Previous Name}
}
\startdata

\cutinhead{Definite Seyfert 2 Galaxies}

2MASS 09444446+0035446 &   9:44:44.46 +00:35:44.6 & 17.9 & 15.15 & 1.52 & 0.1659 & \nodata \\
2MASS 09412757-0020337 &   9:41:27.57 -00:20:33.7 & 16.2 & 15.31 & 1.21 & 0.1487 & \nodata \\
2MASS 09443759+0034107 &   9:44:37.59 +00:34:10.7 & 17.5 & 14.75 & 1.52 & 0.1447 & \nodata \\
2MASS 09410612-0028238 &   9:41:06.12 -00:28:23.8 & 16.4 & 14.81 & 1.47 & 0.1481 & \nodata \\
2MASS 09472633-0005562 &   9:47:26.33 -00:05:56.2 & 17.4 & 14.96 & 1.47 & 0.1260 & \nodata \\
2MASS 09415313+0009185 &   9:41:53.13 +00:09:18.5 & 16.5 & 14.13 & 1.75 & 0.1221 & \nodata \\
2MASS 09422430-0000051 &   9:42:24.30 -00:00:05.1 & 16.1 & 14.16 & 1.80 & 0.1465 & \nodata \\
2MASS 09425917+0031414 &   9:42:59.17 +00:31:41.4 & 16.3 & 14.36 & 1.39 & 0.0633 & \nodata \\
2MASS 09430377+0008076 &   9:43:03.77 +00:08:07.6 & 17.0 & 14.93 & 1.33 & 0.1240 & \nodata \\
2MASS 09452964-0021547 &   9:45:29.64 -00:21:54.7 & 13.2 & 14.00 & 1.24 & 0.0515 & \nodata \\
2MASS 09451196-0007119 &   9:45:11.96 -00:07:11.9 & 12.4 & 13.64 & 1.37 & 0.0306 & \nodata \\
2MASS 09441489+0018082 &   9:44:14.89 +00:18:08.2 & 17.1 & 14.75 & 1.47 & 0.1223 & \nodata \\
2MASS 09443030+0045287 &   9:44:30.30  00:45:28.7 & 17.0 & 14.88 & 1.32 & 0.1237 & \nodata \\
2MASS 12432177+0015370 &  12:43:21.77  00:15:37.0 & 17.1 & 14.86 & 1.38 & 0.1433 & \nodata \\
2MASS 13031198-2447024 &  13:03:11.98 -24:47:02.4 & 17.8 & 15.22 & 1.31 & 0.1252 & \nodata \\
2MASS 12593796-2523148 &  12:59:37.96 -25:23:14.8 & 15.6 & 13.87 & 1.50 & 0.0728 & \nodata \\
2MASS 12595227-2516420 &  12:59:52.27 -25:16:42.0 & 12.8 & 13.01 & 1.36 & 0.0486 & \nodata \\
2MASS 14143610-2546458 &  14:14:36.10 -25:46:45.8 & 16.6 & 15.27 & 1.55 & 0.1653 & \nodata \\
2MASS 14132413-2615549 &  14:13:24.13 -26:15:54.9 & 16.0 & 14.54 & 1.72 & 0.1717 & \nodata \\
2MASS 14140715-2528597 &  14:14:07.15 -25:28:59.7 & 16.2 & 14.66 & 1.33 & 0.1695 & \nodata \\
2MASS 14143799-2520079 &  14:14:37.99 -25:20:07.9 & 17.2 & 14.85 & 1.35 & 0.1391 & \nodata \\
2MASS 14154583-2518245 &  14:15:45.83 -25:18:24.5 & 16.5 & 14.50 & 1.53 & 0.0750 & \nodata \\
2MASS 14155435-2557332 &  14:15:54.35 -25:57:33.2 & 16.8 & 14.73 & 1.33 & 0.1681 & \nodata \\
2MASS 14155854-2544131 &  14:15:58.54 -25:44:13.1 & 16.2 & 14.44 & 1.48 & 0.1697 & \nodata \\

\cutinhead{Probable Seyfert 2 Galaxies}

2MASS 09430751-0002492 &   9:43:07.51 -00:02:49.2 & 17.6 & 15.37 & 1.33 & 0.1247 & \nodata \\
2MASS 09452796+0051041 &   9:45:27.96 +00:51:04.1 & 16.1 & 14.52 & 1.32 & 0.1431 & \nodata \\
2MASS 12445756-0016176 &  12:44:57.56 -00:16:17.6 & 16.8 & 14.50 & 1.30 & 0.1186 & \nodata \\
2MASS 12451295-0040566 &  12:45:12.95 -00:40:56.6 & 15.8 & 14.49 & 1.28 & 0.1043 & \nodata \\
2MASS 12452522-0046579 &  12:45:25.22 -00:46:57.9 & 15.7 & 14.85 & 1.45 & 0.0806 & \nodata \\
2MASS 12595666-2439065 &  12:59:56.66 -24:39:06.5 & 17.4 & 15.24 & 1.23 & 0.1053 & \nodata \\
2MASS 13015692-2533499 &  13:01:56.92 -25:33:49.9 & 17.7 & 15.12 & 1.55 & 0.1859 & \nodata \\
2MASS 12583172-2508040 &  12:58:31.72 -25:08:04.0 & 16.9 & 15.27 & 1.28 & 0.1045 & \nodata \\
2MASS 13005827-2423430 &  13:00:58.27 -24:23:43.0 & 15.3 & 14.27 & 1.24 & 0.0993 & \nodata \\
2MASS 12591093-2446418 &  12:59:10.93 -24:46:41.8 & 15.3 & 13.63 & 1.73 & 0.1125 & \nodata \\
2MASS 12594636-2535568 &  12:59:46.36 -25:35:56.8 & 12.9 & 13.66 & 1.49 & 0.0639 & ARP 1257-251 \\
2MASS 14175559-2538535 &  14:17:55.59 -25:38:53.5 & 17.4 & 14.84 & 1.47 & 0.1209 & \nodata \\
2MASS 14143476-2522301 &  14:14:34.76 -25:22:30.1 & 17.3 & 14.79 & 1.64 & 0.1159 & \nodata \\

\enddata
\end{deluxetable}

\begin{figure}
\plotone{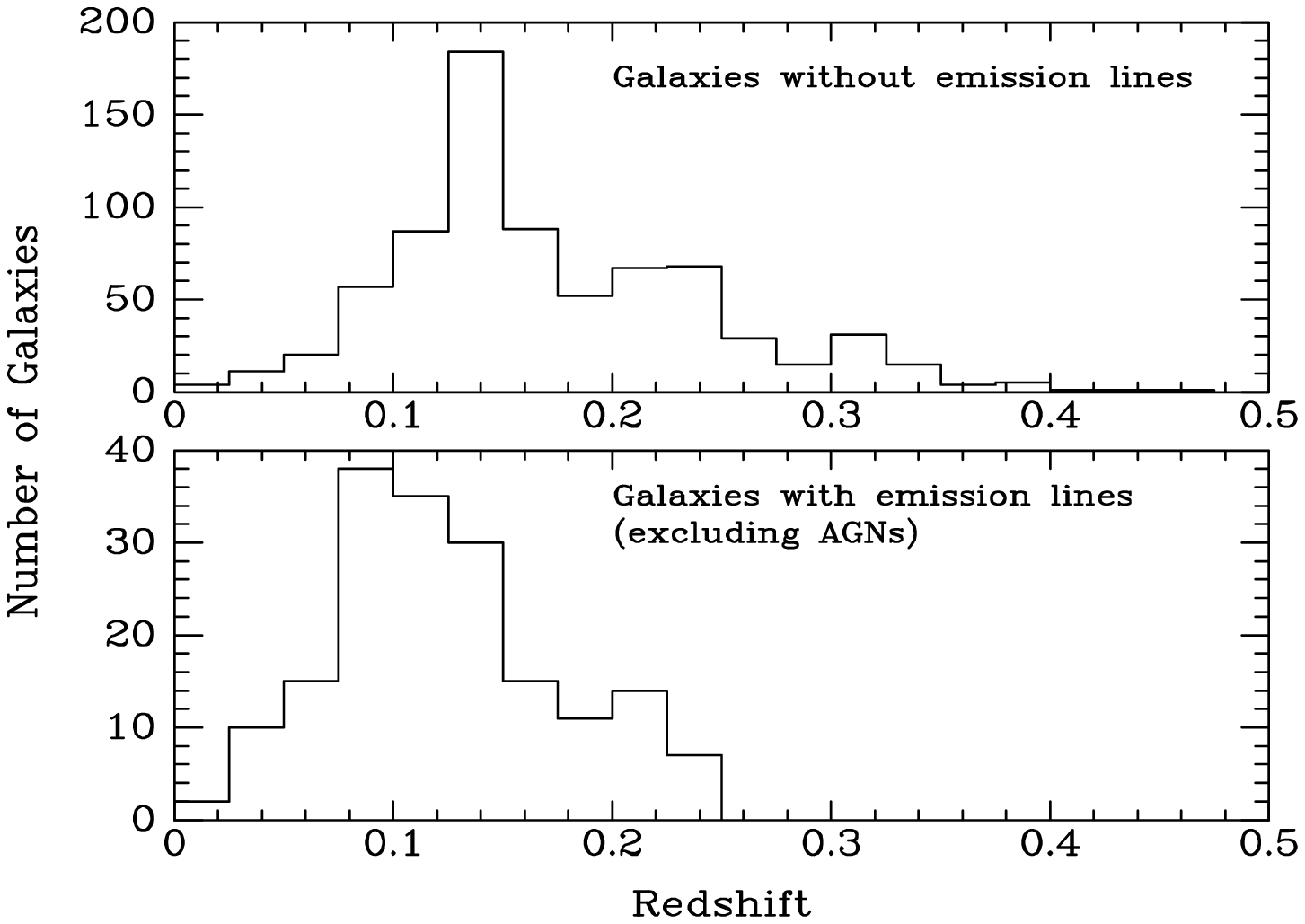}
\caption{
Redshift histogram for galaxies without H$\alpha$ emission (top)
and with H$\alpha$ emission (bottom) down to our equivalent width threshold. 
Known or probable AGN have been
excluded.
\label{gal_zhist}}
\end{figure}

SDSS classifies 91\% of the galaxies without
measurable H$\alpha$ emission as elliptical galaxies (ie. a De Vaucouleurs
profile fits significantly better than an exponential profile). For
emission-line galaxies (excluding AGNs) the fraction is 50\%. The redshift
histograms are shown in Fig~\ref{gal_zhist}. The galaxies without
emission lines are quite strongly clustered: the peak seen at redshift
0.14 is due to one such cluster. The emission-line galaxies lie at a lower
mean redshift than those without emission lines. This is probably because
the emission-line galaxies are late type while those without emission lines
are massive luminous early type galaxies, and hence seen to larger distances.

\begin{figure}
\plotone{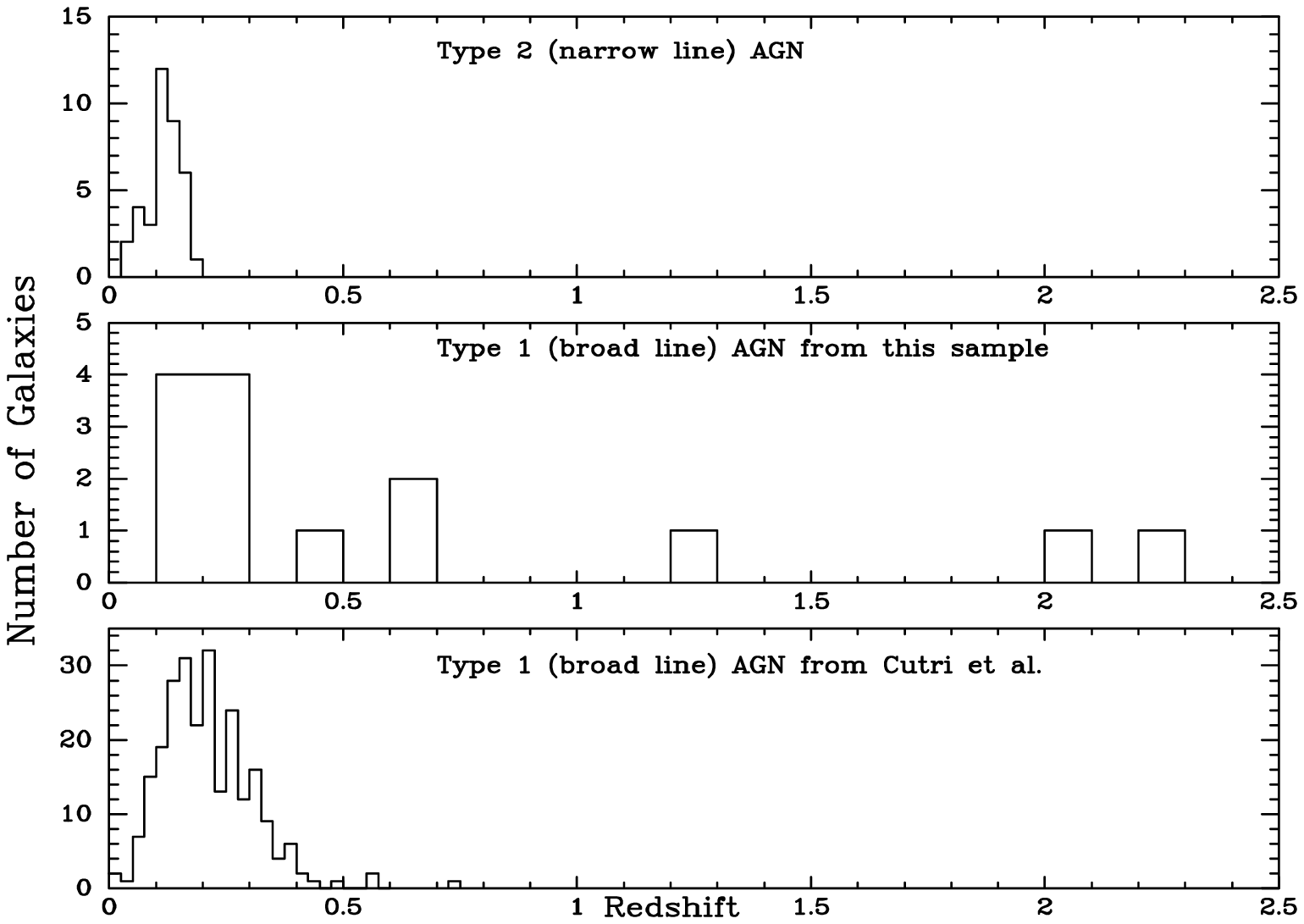}
\caption{
Redshift histogram for Type 2 (narrow-line) AGN (top), Type 1 (broad-line)
AGN (middle) from this sample, and Type 1 AGN with $J-K_s>2$ from the
sample of Cutri et al. (only southern hemisphere AGN with $K_s <14.0$  and
$R<18$ shown).
\label{AGN_zhist}}
\end{figure}

The Type 2 AGNs have a redshift distribution indistinguishable from that of
other emission-line galaxies in the survey (Fig~\ref{AGN_zhist}). Most
Type 1 AGNs also lie at low redshifts, but there is a tail to very high 
redshifts.

\section{Discussion\label{discuss}}

\subsection{The QSO Sample\label{kband}}

We find 12 Type-1 AGNs with $1.2 < J-K_s < 2.0$. Allowing for our
incomplete spectroscopy of the faintest sources, this implies a surface
density of $1.7 \pm 0.5$ Type-1 AGNs per square degree, down to our
selection limits (3-band detection by 2MASS in J, H and K). Many of our
sources, however, are fainter than the nominal completeness limit of the
2MASS survey. We estimated this completeness limit for our fields by comparing
our $K_s$-band galaxy counts against the compilation of \citet{hua01}.
Down to $K_s=15$, our target list appear to be highly complete: the statistical error due 
to our small sample of AGNs is much greater than any error caused by
sample incompleteness. At this limit, and correcting for the unobserved
targets, we are finding
$1.0 \pm 0.3$ Type-1 AGNs per square degree.

Are we missing many AGNs with $J-K_s<1.2$? \citet{bar01} showed that essentially
all low redshift ($z<0.5$) AGNs discovered by other techniques have $J-K_s>1.2$,
and should hence have been found in our survey. At higher redshifts, however,
most AGNs have bluer $J-K_s$ colors. This is probably a $k$-correction
effect: most AGN show a sharp 
rise in flux between rest-frame 1 and 2 microns, perhaps due to hot dust 
emission \citep[eg.][]{san89}, and at redshifts above 0.5, this is redshifted
out of the K-band. Only $\sim 10$\% of high-z ($z>0.5$) AGNs detected by other
techniques have $J-K_s>1.2$, except in the redshift range $2.2 < z < 2.5$,
in which the H$\alpha$ emission line lies within the K-band. This is
consistent with our redshift distribution (Fig~\ref{AGN_zhist}).

Another way to test our completeness is to see whether we recovered
previously know AGNs in our fields. The NASA Extragalactic Database (NED)
lists 99 AGNs in our field, of which only nine meet our magnitude limits.
Only one of the nine has $J-K_s<1.2$. We recovered six of the remaining nine
objects: we did not put fibers on the other two sources.

We can therefore place a lower limit on the surface density of
Type-1 AGN of $1.0 \pm 0.3$ per square degree, down to $K=15$. This
matches the surface density of optically selected AGNs down to $B \sim 18.5$
\citep{mey01}. Given that a typical quasar has
$B-K \sim 3.5$ \citep{fra00}, this suggests that we may be seeing the
same population sampled by optical surveys. This comparison should be
treated with caution, however, as most optically selected QSOs down to
$B=18.5$ lie at redshifts to which we are largely insensitive, and most of
our AGN have such low luminosities that they would be discarded from most
optical samples due to host galaxy contamination (\S~\ref{find}).

\subsection{Could these QSOs be found by conventional techniques?\label{find}}

As Table~\ref{type1} shows, only 57\% of the Type-1 AGNs, and none of the 
Type-2
AGNs had been previously identified as AGNs. All 5 Type-1 AGNs with redshifts
above 0.5 had been previously identified.

\begin{figure}
\plotone{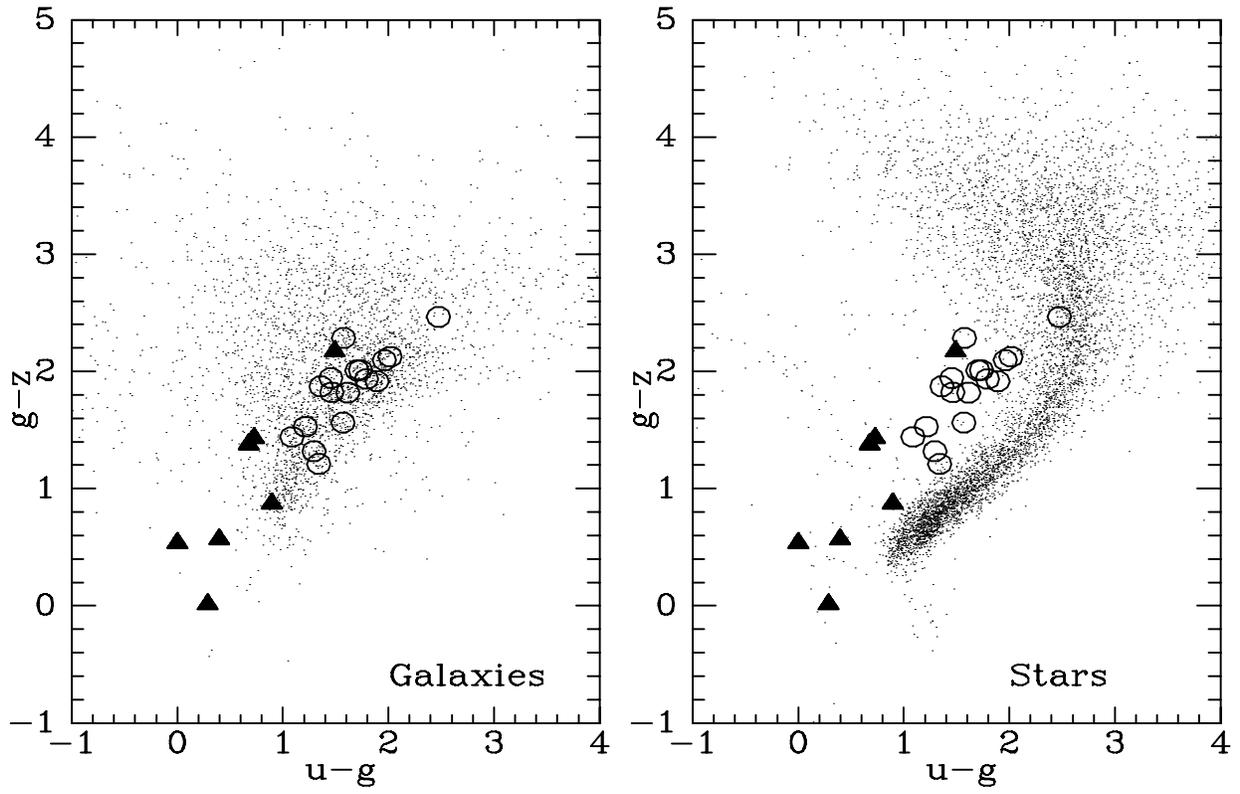}
\caption{
The optical colors of our Type-1 (triangles) and type-2 (circles) AGNs,
compared to all SDSS sources classified as galaxies (left panel) and
stars (right panel).
\label{colourplot}
}
\end{figure}

In Fig~\ref{colourplot}, we compare the optical colors of our AGNs with the
colors of field stars and galaxies. Only sources overlapping with the SDSS
early data release are shown. Our AGNs are clearly separated from the stellar
locus. Half the Type-1 AGNs are also well separated from the galactic locus,
but the other half are not, and the Type-2 AGNs also lie well within the
galactic locus. This explains why so many of our sources were not previously
identified: they are spatially resolved and have galaxy-like optical colors.
Adding near-IR photometry doesn't help: their spectra energy distributions are
indistinguishable from those of inactive galaxies all the way from the $U$ to
the $K_s$ band.

\begin{figure}
\plotone{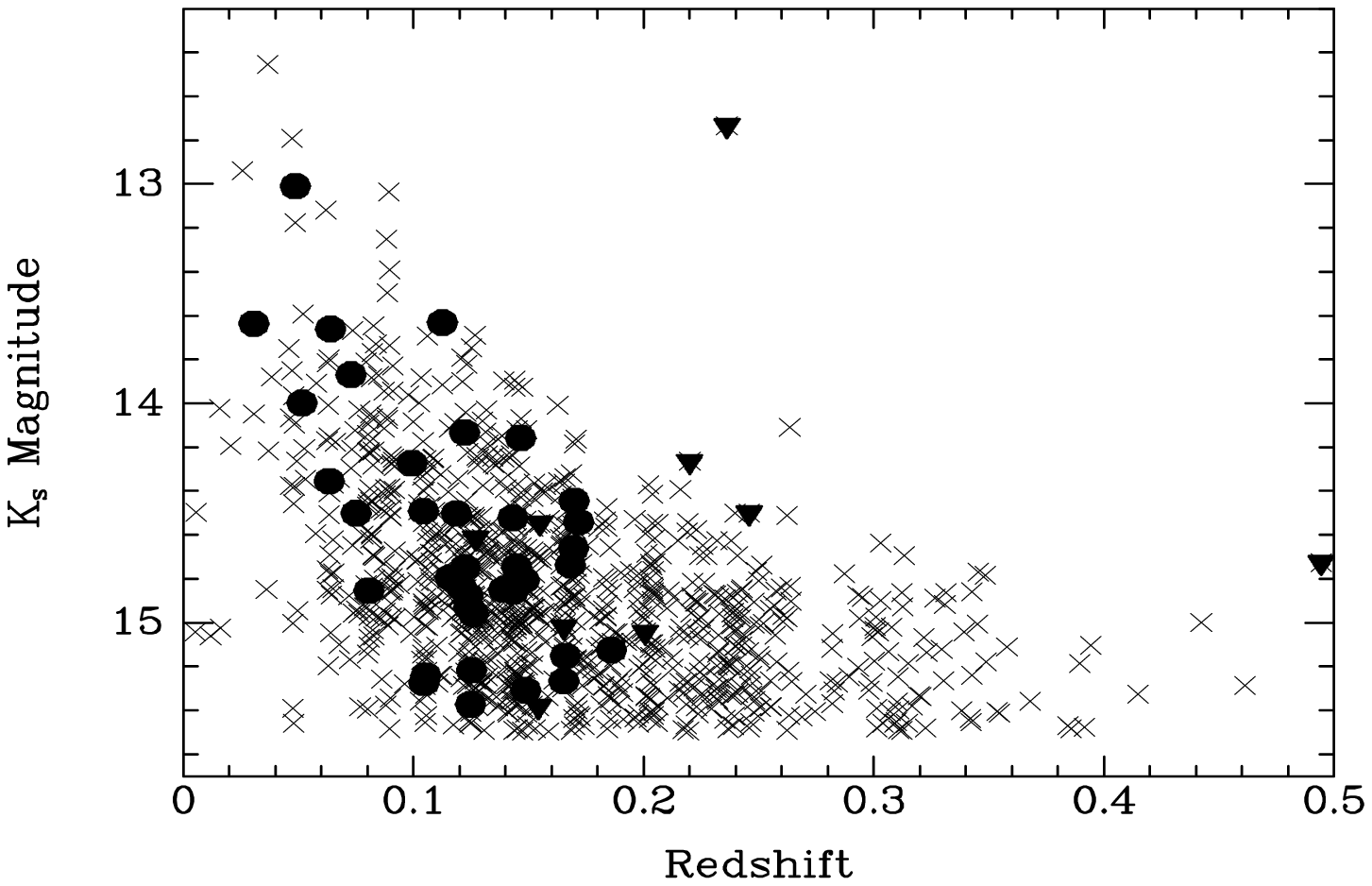}
\caption{
The redshifts and K-band magnitudes of all galaxies in our sample. Type-1 AGN
are shown as filled triangles and Type-2 AGNs are shown as filled circles.
\label{kmag}
}
\end{figure}

Why do the colors of so many of our sources resemble galaxies? The 
spectra of
all AGNs with galaxy-like colors show strong stellar absorption lines, so
their colors are probably dominated by the host galaxy, and not by nuclear
emission. Furthermore, Fig~\ref{kmag}
shows that the $K_s$-band magnitudes of these AGNs are comparable to those
of inactive galaxies at the same redshifts. We thus conclude that the host
galaxy, rather than the nucleus, is dominating the observed continuum flux
at all wavelengths.

While these AGNs would not be identified as such on the basis of their broad-band
colors, their strong broad H$\alpha$ emission lines would make them detectable
in objective prism surveys, and large galaxy
redshift surveys should contain thousands of them.

\subsection{Dusty QSOs?}

Is there a large population of dusty red QSOs? We are unable to determine
whether our low redshift AGNs are dust-reddened, as the host galaxies
dominate their broad-band colors, and as our observations are not
spectrophotometric, we cannot determine the reddening by looking at line
ratios. It is possible that their spectra are dominated by host galaxy light
precisely because they are dusty. Alternatively, their nuclei could be 
intrinsically less luminous, or their host galaxies intrinsically more
luminous than those of optically selected AGNs with the same $K_s$-band 
luminosities.

This leaves the small number of high redshift QSOs. SDSS colors
are available for four of these sources. All four are quite blue: their
mean $g-K_s$ color is 2.86, which is very comparable to that seen in optically
selected QSO samples \citep{fra00}. None are more than 0.7 mag redder than the
mean in $g-K$, corresponding to $A_V=0.8$ (for dust with an optical depth
inversely proportional to wavelength).

We thus see no evidence for a population of dusty red QSOs. Our sample
is, however, too small to place strong constraints. Let us define a red
QSO as one with $g-K_s>3.5$, corresponding to $A_V > 0.8$. 
The fact that none of the four high-z QSOs with SDSS data meet this definition
allows us to say with 95\% confidence that no more than 50\% of QSOs 
down to our magnitude limit are red. As shown in Fig~\ref{dust}, however, 
imposing even a K-band magnitude limit will suppress the numbers of red AGNs
by a factor of $\sim 5$. Our limit is thus a weak one: no more than 80\% of
QSOs can be red. This is quite consistent with the limits derived from radio
surveys \citep{fra01,gre02}.

\subsection{The Fraction of Galaxies with Active Nuclei}

What fraction of the galaxies in our sample contained active nuclei, and
how does this compare to the fraction found in blue galaxy samples?

We are only sensitive to AGNs with H$\alpha$ rest-frame equivalent widths of
greater than 0.4nm, within our aperture (our fibers are 2\arcsec in
diameter, which for the median redshift of the sample (0.15) corresponds to 
a physical diameter of 5kpc). \citet{ho97} showed that we will miss most
AGNs at this equivalent width limit, with our large aperture. In particular, 
we are mainly sensitive to Seyfert-2 galaxies and not to LINERs (low-ionization 
nuclear emission-line regions).

We obtained 1467 usable spectra, of which 330 were stars. Another
169 showed no significant emission or absorption features and are
probably inactive galaxies. The remaining 968 are galaxies for which we
were able to measure secure redshifts. 6\% of these
galaxies had something wrong with their spectra at the wavelength of H$\alpha$.
187 of the remaining galaxies had H$\alpha$ equivalent widths of $>0.4$nm. Of 
these, we were able to spectrally classify 116. We identified 14 Type-1 
AGNs and 35 Type-2 definite or probable AGNs.

There are 71 galaxies with narrow H$\alpha$ lines above our selection
threshold but with the signal-to-noise ratio
too poor to allow classification. If we assume that these galaxies  
have the same relative proportions of Type-2 AGNs and Starburst 
galaxies as the 116 we could classify, then there are 58 Type-2 AGNs in 
the sample.

The sample population in which we could have seen Type-2 AGN activity is
$968+169=1137$ galaxies. We thus estimate that $58/1137 = 5.1 \pm 0.9$\% of
galaxies in our sample were Type-2 AGNs, down to our H$\alpha$ equivalent
width limit (Poisson errors). If none of the unclassifiable emission-line
galaxies were AGNs, which seems unlikely, the fraction would be 
$35/1137 = 3.1$\%. This gives a lower limit on the fraction.
For Type-1 AGNs, the fraction is $1.2 \pm 0.3$\%. Note also that our 35
definite or probable AGNs included several classified on the basis of their
red emission lines only, as discussed in Section~\ref{eline}.

How does this compare with the AGN fraction in blue-selected galaxy samples?
\citet{huc92} only found Seyfert activity (of any type) in 1.3\% of a sample
of 2399 nearby galaxies. Unfortunately, they do not list their equivalent width
threshold, so this number cannot be compared to our figure. \citet{ho97},
however, find that nearly 50\% of nearby blue-selected galaxies are AGNs, 
but they used nuclear spectra and were sensitive to much weaker lines 
than we are.

We defined a sub-sample of the \citet{ho97} sources that would have been 
classified as Seyfert galaxies by our criteria. Firstly, we had to correct for
the different spectroscopic aperture. They typically measured the spectrum
over a region of radius $< 200$ pc, compared to our physical radii of 
$\sim 2500$pc. To see how much difference this typically makes,
we obtained archival CCD images of six Seyfert-2 galaxies 
from their sample, using the NASA Extragalactic Database.
We measured the broad-band optical flux for each galaxy in a 200pc aperture
and a 2500pc aperture. The fluxes in the larger apertures were greater by a 
factor of between 4 and 25.

Thus for one of their galaxies to have made it into our sample (if it were
at the median redshift of our galaxies), it would need a nuclear  
H$\alpha$ equivalent width exceeding 1.6nm ($4 \times 0.4$nm). \citet{ho97} 
also use slightly different diagnostics to identify AGN but this makes little
difference to the final numbers.

We find that $1.5 \pm 0.6$ of their galaxies contained Type-2 AGNs
(meeting our selection criteria) and $1 \pm 0.4$\% contained Type-1 AGNs.
There is thus no significant difference in the fraction of Type-1 AGNs, but
we are finding a significantly higher fraction of Type-2 AGNs.

Why do we find a higher fraction of galaxies with AGNs? Finding an AGN
requires the presence of a black hole, a suitable accretion rate of mass on to
it, gas to be ionized by the nucleus and a dust geometry that both allows this
ionization to take place and allows us to see the resultant narrow line 
emission. Many of these factors could be different in an IR-selected sample.
As black hole masses are correlated with the stellar
masses of the bulge \citep[eg][]{mag98}, this may indicate that our sample
of galaxies contain larger nuclear black holes than those found in blue-selected
samples. On the other hand, blue galaxies typically have more gas and a higher
star formation rate than red ones. Note also that our galaxies lie at
higher redshifts than the \citet{ho97} sample, and are on average more massive
and luminous. We may simply be looking at a tendency for AGNs to be found in
the most massive galaxies.

Could the difference simply be because the near-IR selected galaxies have less
continuum flux at the wavelength of H$\alpha$? Our galaxies have a median
$r-K_s = 2.77$, while blue-selected galaxies at similar magnitude limits have
a median $r-K_s \sim 1.5$. Thus the younger stellar populations in the blue
selected galaxies are increasing the continuum flux per unit stellar mass
by a factor of $\sim 3$. Even allowing for this, the fraction of Seyfert-2
galaxies in \citet{ho97} would only rise to $\sim 3$\% .

This result should be considered tentative. Secure line diagnostics are only
available for around half of the Seyfert-2 population in our sample, and the
comparison with the very different \citet{ho97} sample involves large 
corrections for the different AGN detection thresholds.

\section{Conclusions\label{conclude}}

We have selected a small sample of AGNs in the near-IR, using the brute-force
power of the 2dF spectrograph to minimize selection biases. Perhaps the most
surprising thing about this sample is how similar it looks to conventional
blue-selected AGN samples. While many of our Type-1 AGN would not have been
found by optical techniques, in all cases this seems to be due to host galaxy
contamination. Large galaxy surveys, such as the 2dF Galaxy Redshift Survey
\citep{col01}, are probably the best way to find such AGNs. Our sample
of high redshift QSOs was too small to usefully constrain the population of
dusty red QSOs.

We tentatively conclude that 
the fraction of galaxies in our sample with AGN emission is 
greater than that found in the blue selected galaxy sample of \citet{ho97}.
There are many possible reasons for this difference and
discriminating between them will be difficult.

Finally, we can extrapolate from our data to estimate the number
of active galactic nuclei in the 2MASS point source catalog. There should be 
$\sim$ 50,000 Type-1 AGNs and $\sim$ 200,000 Type-2 AGNs that 
meet our selection criteria.

\acknowledgments

This publication makes use of data products from the Two Micron 
All Sky Survey, which is a joint project of the University of 
Massachusetts and the Infrared Processing and Analysis Center/California 
Institute of Technology, funded by the National Aeronautics and Space 
Administration and the National Science Foundation.

Funding for the creation and distribution of the SDSS Archive has 
been provided by the Alfred P. Sloan Foundation, the Participating 
Institutions, the National Aeronautics and Space Administration, the 
National Science Foundation, the U.S. Department of Energy, the 
Japanese Monbukagakusho, and the Max Planck Society.

The SDSS is managed by the Astrophysical Research Consortium (ARC) for the
Participating Institutions. The Participating Institutions are The University of
Chicago, Fermilab, the Institute for Advanced Study, the Japan Participation Group,
The Johns Hopkins University, Los Alamos National Laboratory, the
Max-Planck-Institute for Astronomy (MPIA), the Max-Planck-Institute for
Astrophysics (MPA), New Mexico State University, University of Pittsburgh,
Princeton University, the United States Naval Observatory, and the University of
Washington.

This research has made use of the NASA/IPAC Extragalactic Database (NED) which 
is operated by the Jet Propulsion Laboratory, California Institute of
Technology, under contract with the National Aeronautics and Space Administration.

\begin{deluxetable}{ccccccccccc}
\tabletypesize{\scriptsize}
\tablewidth{0pt}
\tablecolumns{11}
\tablecaption{All Galaxies with Secure Redshifts \label{allgal}}
\tablehead{
\colhead{Position} &
\colhead{ ~ } &
\colhead{ ~ } &
\colhead{ ~ } &
\colhead{ ~ } &
\colhead{ ~ } &
\multicolumn{5}{c}{Equivalent Width (nm)}
\\
\colhead{(J2000)} & 
\colhead{Redshift} & 
\colhead{Quality} & 
\colhead{Class} & 
\colhead{$K_s$} & 
\colhead{$J-K_s$} &
\colhead{H$\alpha$} &
\colhead{H$\beta$} &
\colhead{[O III]} &
\colhead{N II} &
\colhead{S II} 
}
\startdata
9:40:06.62 -0:05:14.2 & 0.0629 & 5 & sb? & 14.0 & 1.22 &   0.3 & \nodata & \nodata &   0.3 &   0.1 \\
9:40:08.85  0:15:07.8 & 0.0497 & 5 & gem & 14.9 & 1.66 &   1.6 & \nodata & \nodata &   1.1 &   0.6 \\
9:40:09.09 -0:14:41.1 & 0.1724 & 5 & gal & 14.9 & 1.37 & \nodata & \nodata & \nodata & \nodata & \nodata \\
9:40:10.60  0:13:12.6 & 0.0625 & 5 & sb? & 13.8 & 1.40 &   0.5 & \nodata & \nodata &   0.3 &   0.1 \\
9:40:11.62 -0:11:50.5 & 0.2045 & 5 & gal & 15.4 & 1.36 & \nodata & \nodata & \nodata & \nodata & \nodata \\
9:40:14.73 -0:19:46.7 & 0.1256 & 5 & gal & 15.0 & 1.28 & \nodata & \nodata & \nodata & \nodata & \nodata \\
9:40:19.15  0:04:25.4 & 0.0904 & 5 & sb  & 14.9 & 1.72 &   5.0 &   0.7 &   0.4 &   1.7 &   2.0 \\
9:40:20.75  0:23:32.0 & 0.0162 & 4 & gal & 15.0 & 1.27 & \nodata & \nodata & \nodata & \nodata & \nodata \\
9:40:28.31  0:05:23.3 & 0.2468 & 5 & gal & 15.2 & 1.45 & \nodata & \nodata & \nodata & \nodata & \nodata \\
9:40:31.86 -0:28:43.3 & 0.1542 & 5 & qso & 15.4 & 1.39 & \nodata & \nodata & \nodata & \nodata & \nodata \\
\enddata

\tablenotetext{a}{
The complete version of this table is in the electronic edition of
the Journal. The printed edition contains only a sample.}

\end{deluxetable}


\begin{thebibliography}{}

\bibitem[Alexander et al.(2001)]{ale01}
Alexander, D.M., Brandt, W.N., Hornschemeier, A.E., Garmire, G.P.,
Schneider, D.P., Bauer, F.E. \& Griffiths, R. E. 2001, \aj, 122, 2156

\bibitem[Baker \& Hunstead(1995)]{bak95}
Baker, J.C., \& Hunstead, R.W. 1995, \apjl , 452, L95

\bibitem[Barkhouse \& Hall(2001)]{bar01}
Barkhouse, W.A. \& Hall, P.B. 2001, \aj , 121, 2843

\bibitem[Brotherton et al.(1998)]{bro98}
Brotherton, M.S., Tran, H.D., Becker, R.H., Gregg, M.D, 
Laurent-Muehleisen, S.A., \&  White, R.L. 2001, \apj , 546, 775

\bibitem[Colless et al.(2001)]{col01} Colless, M.M. et al. 2001, \mnras ,
328, 1039

\bibitem[Comastri et al.(1995)]{com95}
Comastri, A., Setti, G., Zamorani, G., \& Hasinger, G. 1995, \aap, 296, 1

\bibitem[Courbin et al.(1998)]{cou98}
Courbin, F., Lidman, C., Frye, B.L., Magain, P., Broadhurst, T.J., 
Pahre, M.A., \& Djorgovski, S.G. 1998, \apjl , 499, L119

\bibitem[Croom, Warren \& Glazebrook(2001)]{cro01}
Croom, S.M., Warren, S.J., Glazebrook, K. 2001, \mnras , 328, 150

\bibitem[Cutri et al.(2002)]{cut02} Cutri, R.M., Nelson, B.O.,
Francis, P.J. \& Smith, P. 2002, in AGN Surveys, ASP Conf. Series vol 284,
eds R.F> Green, E. Ye. Kachikian, D.B. Sanders, 127

\bibitem[Dopita et al.(1998)]{dop98}
Dopita, M.A., Heisler, C., Lumsden, S. \& Bailey, J. 1998, \apj,
498, 570

\bibitem[Francis et al.(2001)]{fra01}
Francis, P.J., Drake, C.L.,
 Whiting, M.T., Drinkwater, M.J. \& 
 Webster, R.L. 2001, PASA, 18, 221

\bibitem[Francis, Whiting \& Webster(2000)]{fra00} 
Francis, P.J., Whiting, M.T. \& Webster, R.L. 2000, PASA, 17, 56

\bibitem[Gilli, Salvati \& Hasinger(2001)]{gil01}
Gilli, R., Salvati, M. \& Hasinger, G. 2001, \aap, 366, 407

\bibitem[Gregg et al.(2002)]{gre02}
Gregg, M.D., Lacy, M, White, R.L., Glikman, E., Helfand, D., Becker, R.H.
\& Brotherton, M.S. 2002, \apj, 564, 133

\bibitem[Ho, Filippenko \& Sargent(1997)]{ho97} Ho, L.C., Filippenko, A.V.
\& Sargent, W.L.W. 1997, \apjs , 112, 315

\bibitem[Huang et al.(2001)]{hua01}
Huang, J.-S. 2001, \aap , 368, 787

\bibitem[Huchra \& Burg(1992)]{huc92} Hucra, J. \& Burg, ,R. 1992, \apj ,
393, 90

\bibitem[Jarrett et al.(2000)]{jar00} Jarett, T.H., Chester, T., Cutri, R.,
Schneider, S., Skrutskie, M. \& Huchra, J.P. 2000, \aj , 119, 2498

\bibitem[Kaspi et al.(2000)]{kas00}
Kaspi, S., Smith, P.S.,
Netzer, H., Maoz, D., Jannuzi, B.T. \&  Giveon, U. 2000, \apj , 533, 631

\bibitem[Kewley et al.(2001)]{kew01} Kewley, L., Heisler, C., Dopita, M. \&
Lumsden, S. 2001, \apjs , 132, 37

\bibitem[Lewis et al.(2002)]{lew02} Lewis, I.J. et al. 2002, \mnras ,
333, 279

\bibitem[Low et al.(1988)]{low88} 
Low, F.J., Cutri, R.M., Huchra, J.P. \& Kleinmann, S.G. 1988,
\apj, 327, L41

\bibitem[Magorrian, J. et al.(1998)]{mag98} Magorrian, J. et al. 1998,
\aj , 115, 2285

\bibitem[Malhotra, Rhoads \& Turner(1997)]{mal97}
Malhotra, S., Rhoads, J. E., \& Turner, E. L. 1997, \mnras , 288, 138 

\bibitem[Matute et al.(2002)]{mat02}
Matute, I. et al. 2002, \mnras , 332, 11

\bibitem[McDowell et al.(1989)]{mcd89}
McDowell, J.C.; Elvis, M.,
 Wilkes, B.J., Willner, S.P.,
 Oey, M.S., Polomski, E.
 Bechtold, J. \& Green, R.F. 1989, \apjl , 345, L13

\bibitem[Meyer et al.(2001)]{mey01}
Meyer, M.J., Drinkwater, M.J., Phillips, S. \& Couch, W.J. 2001 \mnras, 
324, 343

\bibitem[Mushotzky et al.(2000)]{mus00}
Mushotzky, R.F., Cowie, L.L., Barger, A.J. \& Arnaud, K.A. 2000, Nature, 
404, 459

\bibitem[Richards et al.(2002)]{ric02}
Richards, G.T. et al. 2002, \aj , 123, 2945

\bibitem[Sanders et al.(1989)]{san89}
Sanders, D.B., Phinney, E.S., Neugebauer, G., Soifer, B.T. \& Matthews,
K. 1989, \apj , 347, 29

\bibitem[Schlegel, Finkbeiner \& Davis(1998)]{sch98}
Schlegel, D.J., Finkbeiner, D.P. \& Davis, M.1998,  \apj, 500, 525 

\bibitem[Skrutskie et al.(1997)]{skr97} Skrutskie, M.F. et al. 1997, in 
``The Impact of Large Scale 
Near-IR Sky Surveys'', eds. F. Garzon et al. (Kluwer, Netherlands), 25

\bibitem[Smith et al.(2002)]{smi02}
Smith, P.S., Schmidt, G.D., Hines, D.C., Cutri, R.M. \& Nelson, B.O.
2002, \apj, 569, 23

\bibitem[Stoughton et al.(2002)]{sto02} Stoughton, C. et al. 2002, \aj ,
123, 485

\bibitem[Warren, Hewett \& Foltz(2000)]{war00}
Warren, S.J., Hewett, P.C. \& Foltz, C.B., 2000, \mnras, 312, 827

\bibitem[Webster et al.(1995)]{web95} 
Webster, R.L., Francis, P.J., Peterson, B.A., Drinkwater, M.J.
\& Masci, F.J. 1995, Nature, 375, 469

\bibitem[Wilkes et al.(2002)]{wil02}
Wilkes, B.J., Schmidt, G.D., Cutri, R.M., Ghosh, H., Hines, D.C.,
 Nelson, B. \&  Smith, P.S. 2002, \apj , 564, 65

\bibitem[Whiting, Webster \& Francis(2001)]{whi01}
Whiting, M.T., Webster, R.L. \& Francis, P.J. 2001, \mnras 323, 718 


\end{thebibliography}
\end{document}